\documentclass[pre,aps,twocolumn,showpacs]{revtex4}  %showpacs

\usepackage{graphicx}
\usepackage{amsmath}

\begin{document}

\title{
Fragility in p-spin models
}
\author{
G.~Parisi$^{1,2}$, G.~Ruocco$^{1}$ and 
F.~Zamponi$^{1}$\footnote{e-mail: francesco.zamponi@phys.uniroma1.it}
        }
\affiliation{
         $^1${Dipartimento di Fisica and INFM, Universit\`a di Roma
         {\em La Sapienza}, P. A. Moro 2, 00185 Roma, Italy}
        }
\affiliation{
        $^2${INFM - Center for Statistical Mechanics and Complexity,
        Universit\`a di Roma {\em La Sapienza}, P. A. Moro 2, 00185 Roma, Italy}
        }

\date{\today}
\begin{abstract}
We investigate the relation between fragility and phase space properties - such as the
distribution of states - in the mean field p-spin model, a solvable model that has been
frequently used in studies of the glass transition.
By direct computation of all the relevant quantities, we find that: {\it i)} the recently
observed correlation between fragility and vibrational properties at low temperature
is present in this model;
{\it ii)} the total number of states is a decreasing function of fragility,
at variance of what is currently believed.
We explain these findings by taking into account the contribution to fragility coming from
the transition paths between different states.
Finally, we propose a geometric picture of the phase space that explains the correlation between 
properties of the transition paths, distribution of states and their vibrational properties.
However, our analysis may not apply to strong systems where
inflection points in the
configurational entropy as a function of the temperature are found.
\end{abstract}
\maketitle

%%%%%%%%%%%%%%%  TEXT  %%%%%%%%%%%%%%%%

\section{Introduction}

The glass-forming materials are characterized by a huge variation
of their transport properties (viscosity, mobility, diffusivity)
and dynamical (relaxation times) properties, upon supercooling.
As an example, the viscosity in the whole "liquid" range, if
crystallization is avoided, increases of about 17 orders of
magnitude by decreasing temperature before the system falls in
the glassy state. The transition to the latter is conventionally
fixed at the temperature, $T_g$, where the viscosity reaches a
value of $10^{13}$ poise. Different systems show different temperature behavior
of the viscosity, and they have been classified accordingly.
Fragility is an index measuring the steepness of the viscosity as
a function of the temperature on approaching the glassy state:
``fragile'' systems (high value of the fragility index) are
characterized by a super-Arrhenius behavior of the viscosity,
that increase very fast and -if extrapolated below $T_g$- seems
to diverge at a finite temperature $T_K$. In ``strong'' systems
(low value of the fragility index), on the contrary, the
viscosity increase is less dramatic and follows an Arrhenius law,
apparently diverging only at zero temperature.

The identification of the microscopic details that, in a given
glass former, determine the temperature dependence of the
viscosity, and thus the value of the fragility, is a long standing
issue in the physics of supercooled liquids and glassy state.
Large numerical and theoretical effort have been devoted to the
attempt to relate the fragility to the specific interparticle
interactions (e.~g. strong glasses are often characterized by
highly directional covalent bonds, while the fragile one have more
or less isotropic interactions). More recently, the attention has
been focused on the possible relation existing between the
fragility and the features of the Potential Energy Landscape
(PEL), more specifically the energy distribution of the minima of
the PEL and the properties of the basin of attractions of such a
minima. With this respect, a key point is the existence of a
relation between viscosity (or relaxation times) and the
configurational entropy $\Sigma(T)$ (i.~e., the number of basins
populated at given temperature), namely the Adam-Gibbs
relation:
\begin{equation}
\label{AG}
\eta(T) = \eta_\infty \exp \left( \frac{\cal E}{T \Sigma(T)} \right)
\end{equation}
This relation has been extensively tested against the experimental
result, and it is now commonly accepted as "correct". Despite its
success in describing both numerical and experimental data, the
Adam-Gibbs relation has not been still derived in a clear way from
microscopic models. This leaves unsolved the question of the
microscopic interpretation of the parameter ${\cal E}$,
that is usually believed to be
related to the properties of the transition paths between
different minima of the potential energy, such as the height of
the barriers or the connectivity of the minima. By using the
Adam-Gibbs relation, one could expect to relate fragility to the
properties of $\Sigma(T)$, i.e., to the distribution of basins in
the phase space of the system. However, this possibility is
frustrated by the lack of knowledge on the parameter ${\cal E}$.
Indeed, once a model for $\Sigma(T)$ has been chosen, one can obtain the whole
range of experimentally observed fragilities by varying 
${\cal E}$ (see Ref.~\cite{noifrag} and references therein).
More specifically, in Ref.~\cite{noifrag} it was observed that for
a large class of models for $\Sigma(T)$ - where $\Sigma(T)$ is a
concave function of $T$ that vanishes at a given temperature
$T_K$ and assumes its maximum $\Sigma^*$ at high temperature
(``Gaussian-like models'') - the relevant parameter that actually
determines the fragility is
\begin{equation}
\label{Ddef} D=\frac{\cal E}{T_K \Sigma^*} \ .
\end{equation}
Thus, fragility appears to be determined by the ratio between
${\cal E}$ (measured in units of $k_B T_K$) and the total number
of states $\Sigma^*/k_B$; it is related to both the distribution
of minima (through $\Sigma^*$) and the characteristic of the
transition path between them (through $\cal E$). The relation
between fragility and phase space properties can be even more
complicated, in those cases where the function $\Sigma(T)$ does
not belong to the Gaussian class.

The relevance of the concept of fragility also relies on the
correlations that have been found between this index and other
properties of glass-forming liquids. Examples of these
correlations are the specific heat jump at $T_g$ (thermodynamic
fragility) \cite{Ang1}, the degree of stretching in the
non-exponential decay of the correlation functions in the liquid
close to $T_g$ \cite{Ngai}, the visibility of the Boson peak at
the glass transition temperature \cite{Sokolov}, or the temperature
behavior of the shear elastic modulus in the supercooled liquid state
\cite{Dyre}. More recently a striking correlation between fragility
and the vibrational properties of the glass at low temperatures
has been found \cite{TS}. Specifically, by examining the dynamic
structure factors of different glass former well below $T_g$, it
has been found that the fragility of the corresponding liquid is
proportional to the rate of change of the non-ergodicity factor in
the $T\rightarrow 0$ limit. Being the latter quantity fully
determined by the (harmonic) vibrational properties (eigenmodes of
the disordered structure), this finding implies the existence of a
deep relation between three features of the PEL: the energy of the
minima, the transition paths between them (that together determine
the fragility) and the Hessian matrix, evaluated at the minima
themselves, that fixes the vibrational properties.

With the aim to elucidate the existence of this unexpected
correlation between energy, curvature and transition paths in the
minima of the PEL we selected a solvable model of ``glass'', where
i) the distribution of minima is ``Gaussian-like'', ii) the
vibrational properties of the minima can be determined, and iii)
the transition path between different minima can be evaluated and
characterized by an energy parameter. More specifically, we
investigate the mean field p-spin model (in both its spherical and
Ising spin version), a model that share with the structural glasses
many aspects of the glass transition phenomenology, and that is
known to have a Gaussian-like distribution of states. Our goal is
twofold: {\it i)} we aim to verify if the analysis reported in
\cite{noifrag} is indeed correct in some microscopic model, i.e.,
if one can obtain a wide range of fragilities in a Gaussian-like
model by varying the parameter $\cal E$, and {\it ii)} to check
whether one can explain the correlation between fragility of the
liquid and the vibrational properties of its glass found in
\cite{TS} by studying the geometry of the phase space.
The latter point could allow us to shed light on
the origin of the correlation between
number of minima, their vibrational properties and the property of
the transition path between them.

The paper is organized as follows: in Sec.~\ref{sec:definitions}
we define the relevant quantities in the case of the mean field
p-spin model; in Sec.~\ref{sec:sferico} we compute them for
the spherical p-spin model, and in Sec.~\ref{sec:ising} for the
Ising p-spin model; in Sec.~\ref{correlazioni} we discuss the
relation between fragility and phase space geometry in these models,
and compare our result with experimental data. Finally, we draw
the conclusions.

\section{Definition of the relevant observables}
\label{sec:definitions}

\noindent
The quantities we wish to compute are (we will set $k_B=1$ in the following):
\begin{equation}
\nonumber
\begin{array}{ll}
T_K & \text{Thermodynamical transition temperature} \\
T_g & \text{Glass transition temperature} \\
T_d & \text{Dynamical transition temperature} \\
\Sigma(T_g) & \text{Complexity at $T_g$} \\
m(T_g) & \text{Fragility} \\
\alpha(T_g) & \text{``Volume'' of the equilibrium states at $T_g$} \\
{\cal E}(T_g) & \text{``Barrier height'' at $T_g$}
\end{array}
\end{equation}
First, we have to identify the proper definition of these
quantities in a mean field model. The main problem is that in a
mean field model the glass transition temperature $T_g$ is not
a well defined quantity. Indeed, the relaxation time of the system is known
to diverge - as a power-law - when the temperature approaches the
dynamical transition temperature $T_d$, that corresponds to the
usual Mode-Coupling temperature $T_{MCT}$. The crossover from a
power-law behavior of the relaxation time to an Arrhenius-like
behavior, observed around $T_{MCT}$ in finite dimensional
systems, is due to the activated processes becoming relevant;
these processes are absent in mean field systems, and the
crossover at $T_{MCT}$ becomes a true dynamical transition at
$T_d$ \cite{LeticiaLH}. To overcome this problem, we will give an
estimate of the height ${\cal E}(T)$ of the barrier that the
system must pass through in order to escape from a metastable
state at a given temperature $T$. Thus, we will make use of a
``fictitious'' Adam-Gibbs relation,
\begin{equation}
\eta(T)=\eta_\infty \exp \left( \frac{{\cal E}(T)}{T \Sigma(T)}
\right) \ ,
\end{equation}
and define $T_g$ by
$\eta(T_g)/\eta_\infty =$ const, or, equivalently, by
\begin{equation}
\frac{{\cal E}(T_g)}{T_g \Sigma(T_g)} =
\text{const} \ .
\end{equation} 
Note that in this paper we will
not distinguish between the ``complexity'' (or ``configurational
entropy'') $\Sigma(T)$, that can be calculated in mean field
models, and the ``excess entropy'' measured in the experiments:
indeed, they behave in a similar way in a wide
class of systems \cite{corezzi}. Obviously, the quantity
$\eta(T)$ has no dynamical meaning in a mean field context, but
it provides an useful definition of $T_g$ that hopefully
coincides with the usual one in finite dimension. It will turn
out that our analysis is not strictly dependent on this
definition of $T_g$, the behavior of the various quantities at $T_g$
being representative, as we will see, of a general trend observed
at all temperatures $T\in [T_K,T_d]$ by varying $p$.

\subsection{Two-replica potential}

\begin{figure}[t]
\centering
%\vspace{.05cm}
\includegraphics[width=.45\textwidth,angle=0]{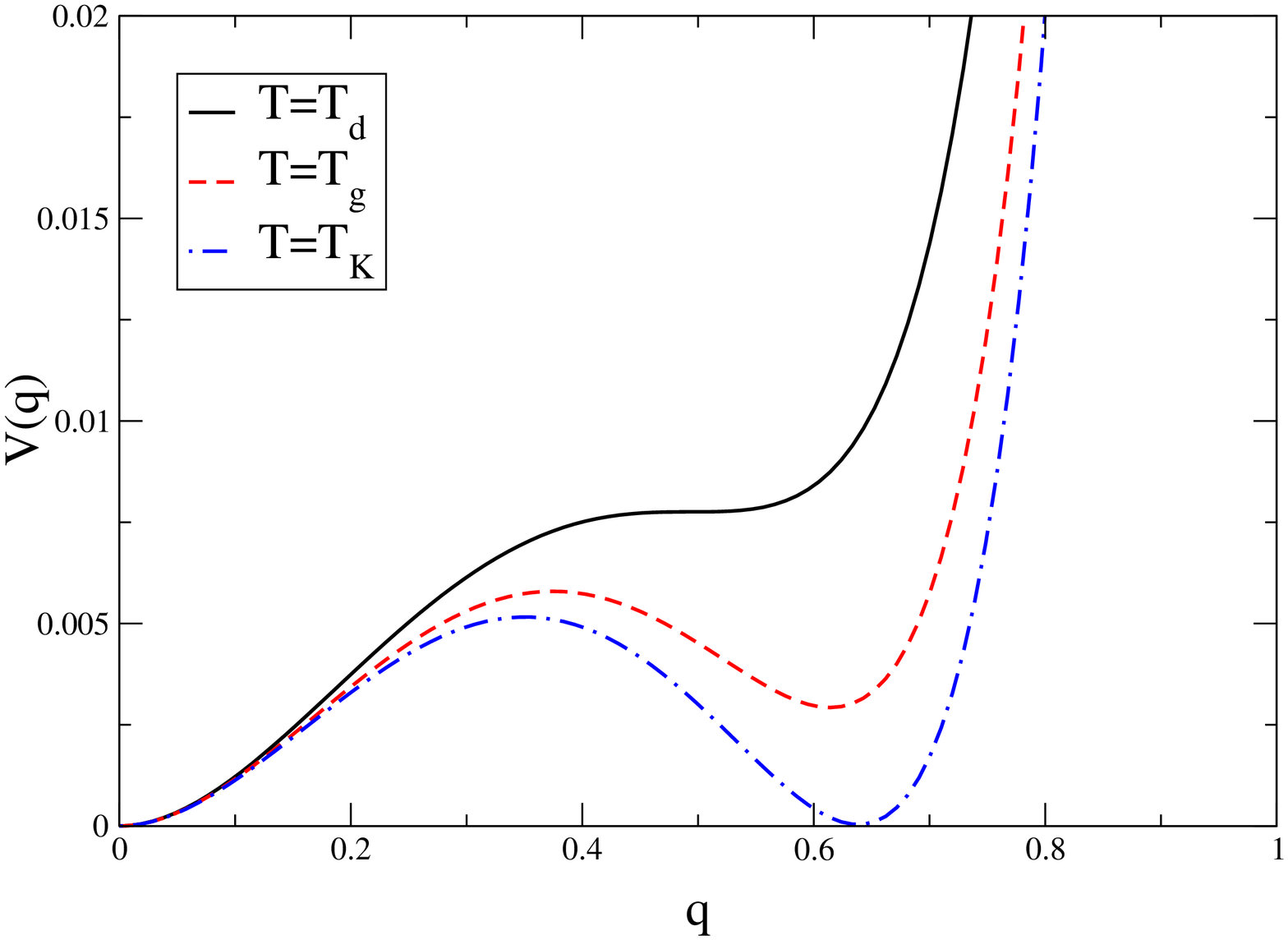}
\caption{The two-replica potential for $T$$\in$$[T_K,T_d]$ in the spherical p-spin model.}
\label{fig_1}
\end{figure}

The easiest way to provide a definition of all the above quantities is to consider the
two-replica potential approach introduced and discussed in \cite{FP1,FP2}.
This function, $V(q,T)$, can be interpreted as the free-energy cost paid to keep two
copies of the system at a fixed overlap~$q$,
\begin{equation}
V(q,T)=F_q(T) - F_0(T) \ ;
\end{equation}
here $F_q(T)$ is the free energy of two copies of the system when constrained to have fixed
overlap $q$, and $F_0(T)=2F(T)$ is the free energy of two independent copies of the system.

Its qualitative behavior is shown in Fig.~\ref{fig_1} for the 1RSB mean field p-spin spherical model:
for $T>T_d$ it is a convex function of $q$ with only one minimum at $q=0$.
At the dynamical transition temperature $T_d$ a secondary minimum starts to develop at finite $q$.
On lowering the temperature below $T_d$, the value of $V$ at the minimum
decreases and vanishes at the thermodynamical transition temperature $T_K$.

From the potential $V(q,T)$ one can extract information about the complexity $\Sigma(T)$ and the
barrier height ${\cal E}(T)$.
Indeed, it is well known that for $T_K<T<T_d$ the phase
space of 1RSB models is disconnected in an exponentially high number of metastable states.
The Gibbs equilibrium state is a superposition of a subset of these states (``equilibrium states'')
having a defined self-overlap $q(T)$;
however, the probability of finding - at equilibrium - two independent
copies of the system in the same state is zero \cite{CGP1}.
Different states have zero overlap, we therefore expect
that the stable phase of the two copies of the system - i.e., the one for which $V(q)$ is minimum -
is at $q=0$. This is indeed the case as one can see from Fig.~\ref{fig_1}.
The secondary minimum at $q\neq 0$ can then be interpreted
as a metastable state for the two coupled systems, that corresponds to the situation where
both systems are in the same state with self overlap $q(T)$.
Thus, the value $q_{min}(T)$ where $V(q,T)$ has a secondary minimum can be interpreted as
the self-overlap of the equilibrium states at temperature $T$.

The free energy of the system for $T_K<T<T_d$ can be written as
\begin{equation}
F(T) = f(T) - T \Sigma(T) \ ,
\end{equation}
where $f(T)$ is the free energy of a single equilibrium state at temperature $T$ and
$\Sigma(T)$ is the complexity, i.e. the logarithm of the number of equilibrium states.
The free energy of two independent copies of the system is $F_0(T) = 2 F(T)=2f(T) - 2T \Sigma(T)$,
while the free energy of two copies constrained to be in the same equilibrium state is given by
\begin{equation}
F_{q_{min}}(T) = 2 f(T) - T \Sigma(T) \ .
\end{equation}
Thus,
\begin{equation}
V(q_{min},T)=F_{q_{min}}(T)-F_0(T)=T\Sigma(T) \ ,
\end{equation}
and the equilibrium complexity $\Sigma(T)$ can be deduced from the function $V(q,T)$.

The difference between the value of $V$ at the maximum and the value of $V$
at the minimum can be interpreted as the height of a ``barrier'' that the
two coupled systems have to overcome to escape from the situation where they are
constrained to be in the same state. Thus, we can define the
``barrier height'' ${\cal E}(T) = V(q_{max},T)-V(q_{min},T)$.
Note that a system-dependent proportionality factor is needed in order to account for the
cooperativity of the process of escaping from a state: indeed, $V(q)$ is the free energy
per spin, while an unknown number of spin can be involved in the escaping process.
Therefore, the ${\cal E}(T)$ defined above is an estimate of the barrier height up to an unknown
(system-dependent) proportionality factor.

\subsection{Temperatures}

The thermodynamical transition temperature $T_K$ is defined as the temperature where
the complexity vanishes: $\Sigma(T_K)=0$.
Then, at $T_K$ the value of $V$ at the secondary minimum becomes equal to zero
(see Fig.~\ref{fig_1}).
The dynamical transition temperature $T_d$ is the temperature at which the metastable
minimum first appears.
We now provide a definition of ``glass transition temperature'' $T_g$.
As we discussed at the beginning of this section,
using the Adam-Gibbs relation, the usual definition of glass
transition temperature turns out to be
\begin{equation}
\label{defTg}
\frac{{\cal E}(T_g)}{T_g\Sigma(T_g)} = {\cal C} \ .
\end{equation}
The value of the constant ${\cal C}$ is arbitrary.
Taking into account the fact that in the considered models ${\cal E}$ is defined
up to a proportionality factor, we can fix the value of the constant in order to
obtain reasonable (with respect to experiments) values for the different quantities
we want to study, fragility in particular.
Different choices of the constant change only quantitatively
the results, while the qualitative picture stays the same.

\subsection{Complexity, barrier heights and fragility}

Given a definition of $T_g$, the complexity at $T_g$ is simply $\Sigma(T_g)$
and the barrier height ${\cal E}(T_g)$: clearly, these two quantities are related
by Eq.~\ref{defTg}.
Knowing the complexity as a function of the temperature,
we can define the fragility as:
\begin{equation}
\label{fragilita}
m(T_g)=1+T_g \frac{\Sigma'(T_g)}{\Sigma(T_g)} \ .
\end{equation}
The latter definition is very useful in a mean field context as - once a definition of
$T_g$ has been chosen - it involves only the complexity,
that is a well-defined quantity in mean field models. It is equivalent to the usual Angell
definition of fragility if $\eta_\infty =$const, and the Adam-Gibbs relation is assumed
to be valid \cite{noifrag}. This definition of fragility has been shown to be related to
the one usually considered in experiments in Ref.~\cite{Ang1}.

\subsection{Volume of the states}

As we discussed in the Introduction, in Ref.~\cite {TS}
fragility has been shown to be correlated with an index
related to the volume of the states populated at equilibrium around $T_g$.
More precisely, in \cite{TS} this index has been defined as
\begin{equation}
\label{alphaTS}
\alpha(T_g) = \lim_{k \rightarrow 0} \left. \frac{d [f_k(T)]^{-1}}{d (T/T_g)} \right|_{T=0} \ ,
\end{equation}
where $f_k(T)$ is the non-ergodicity factor extracted from the dynamic structure factor
$S(k,\omega)$ at a given wave vector $k$. From Fig.~2 of Ref.~\cite{TS}, we see that the
possibility of classifying the considered systems in term of $\alpha$ - given by
Eq.~\ref{alphaTS} - relies on the observation that the curves of $f_k$ as a function of $T/T_g$
for different systems do not intersect (the same observation, that holds for
$\log \eta(T)$ as a function of $T_g/T$, is the basis of the definition of fragility).
Therefore, the index $\alpha$ defined in \cite{TS} can be replaced by other equivalent
- by equivalent we mean positively correlated -
definitions (like the definition of $F_{1/2}$ as a ``fragility index'' \cite{Ang1}).
An useful equivalent definition of $\alpha$ is
\begin{equation}
\alpha(T_g) = \lim_{k \rightarrow 0} \Big[ 1 - f_k(T_g) \Big] \ .
\end{equation}
As one can easily check observing Fig.~2 of Ref.~\cite{TS}, this definition is equivalent
to Eq.~\ref{alphaTS} if the curves $f_k(T)$ do not intersect.

The quantity $f_k(T)$ (in the low-$k$ limit) can be identified in the considered models with
the self-overlap of the states:
this identification come from the observation that both quantities represent the {\it plateau}
of a relevant correlation function.
Thus, we will define
\begin{equation}
\label{alphaGP}
\alpha(T_g) = 1 - q(T_g) \ ,
\end{equation}
where $q(T_g)$ is the self-overlap of the equilibrium states at $T_g$, i.e., the value of
$q$ where $V(q,T)$ has the secondary minimum at $T=T_g$ (see Fig.~\ref{fig_1}).

As the self-overlap of the states is related to their volume in phase space (high overlap
corresponding to small states), a small value of $\alpha$ corresponds to
small-volume states, while a big value of $\alpha$ corresponds to large-volume states.
In this sense, $\alpha(T_g)$ will be called ``volume of the equilibrium states at $T_g$''.
Note that a similar identification has been discussed in Ref.~\cite{TS}: indeed, from
Eq.~7 of Ref.~\cite{TS} (note that due to a typing error the power $-1$ has to be disregarded)
one can see that $\alpha$ is related to the curvatures of the minima
of the potential (in the harmonic approximation), and that small curvatures (large volume)
correspond to large $\alpha$, while high curvatures (small volume) correspond to small $\alpha$.
This is consistent with the equivalence of the definition of $\alpha$ given in Ref.~\cite{TS}
and the one adopted here.

\subsection{Summary of the definitions}

To conclude this section, we give a short summary of all the definition we discussed.
We will call $q_{min}(T)$ the value of $q$ where $V(q,T)$ has the secondary minimum,
and $q_{max}(T)$ the value of $q$ where $V(q,T)$ has a maximum. Then, we define:
\begin{equation}
\nonumber
\begin{array}{lcl}
\Sigma(T) & = & V(q_{min}(T),T)/T \\
{\cal E}(T) & = &  V(q_{max}(T),T)-V(q_{min}(T),T) \\
T_K & : & \Sigma(T_K)=0 \\
T_g & : & \frac{{\cal E}(T_g)}{T_g \Sigma(T_g)} = {\cal C} \\
T_d & : & q_{max}(T_d)=q_{min}(T_d) \\
m(T_g) & = & 1+T_g \frac{\Sigma'(T_g)}{\Sigma(T_g)} \\
\alpha(T_g) & = & 1- q_{min}(T_g)\\
\end{array}
\end{equation}
The constant ${\cal C}$ will be chosen in order for the fragility to be in the
experimentally observed range.

\section{Spherical p-spin model}
\label{sec:sferico}

\begin{figure}[t]
\centering
%\vspace{.05cm}
\includegraphics[width=.45\textwidth,angle=0]{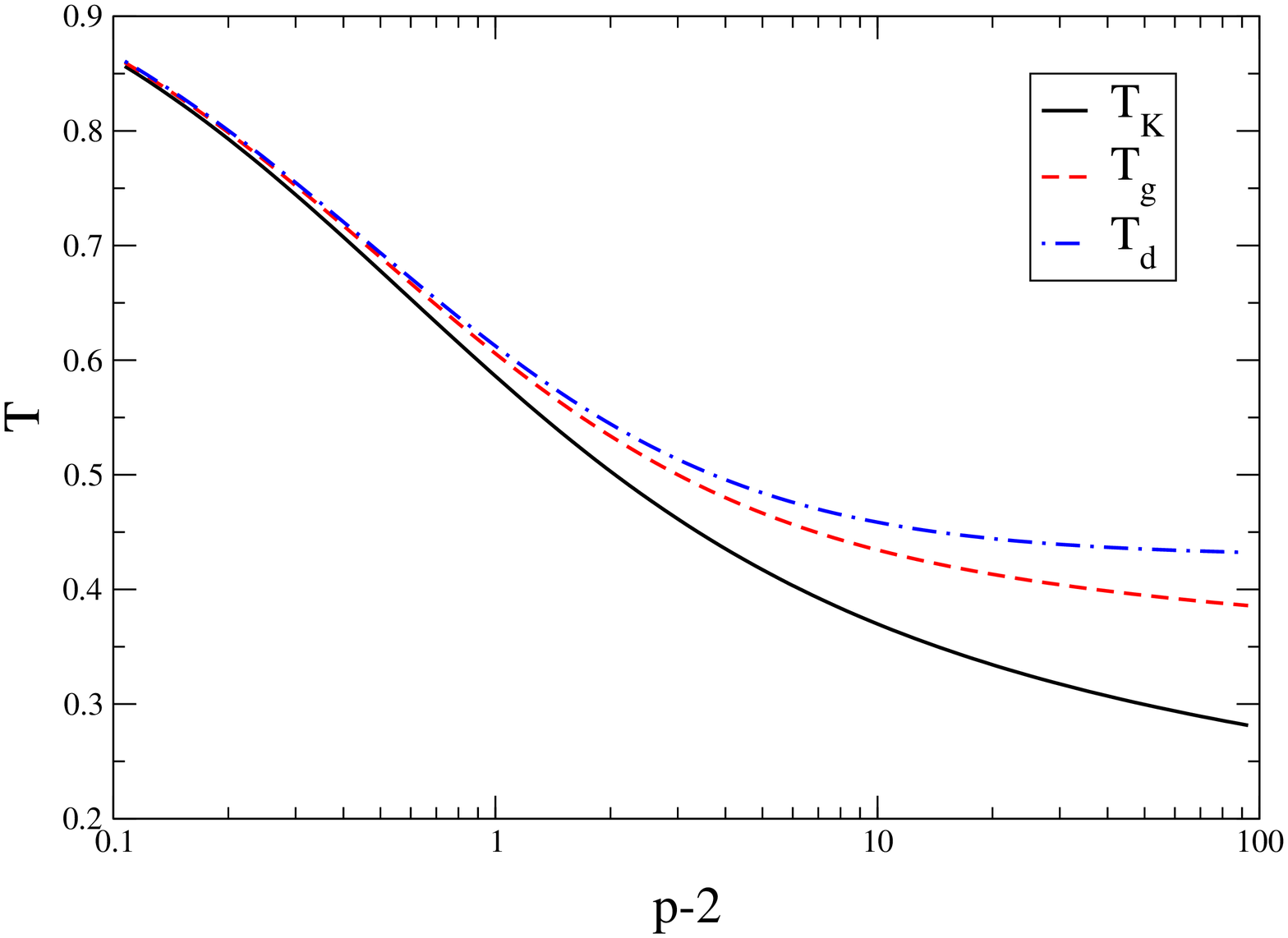}
\caption{Thermodynamic transition temperature $T_K$, glass transition temperature $T_g$
and dynamical transition temperature $T_d$ for the p-spin spherical model as a function
of $p-2$.}
\label{fig_2}
\end{figure}

In this section, we will compute explicitly all the previously defined quantities in the
spherical p-spin model.
The model is defined by the Hamiltonian
\begin{equation}
\label{Hpspin}
H_p = -\sum_{(i_1,\cdots,i_p)} J_{i_1,\cdots,i_p} \sigma_{i_1} \cdots \sigma_{i_p} \ ,
\end{equation}
where $\sigma_i$ are real variables subject to a spherical constraint $\sum_i \sigma^2_i =N$,
and $ J_{i_1,\cdots,i_p}$ are quenched random Gaussian variables with zero mean and variance
$p!/(2 N^{p-1})$.
This simple model has been successfully used for studies of the glass transition
\cite{CS,LeticiaLH}. It is a ``Gaussian-like''
model, in the sense that its complexity - even if the distribution of states in not 
exactly Gaussian -
is known to be a concave function of the temperature,
that vanish at $T_K$ and assumes its maximum at $T_d$, without any inflection point
in between~\cite{CS2}.

The expression for $V(q,T)$ in the p-spin spherical model has been computed
in Refs.~\cite{FP1,BFP}.
However, a simplified expression can be used when the value of
$V(q,T)$ {\it on its stationary points} is considered (see Appendix~\ref{appVq}):
\begin{equation}
\label{Vq}
V(q,T)=-\frac{\beta}{4} q^p - \frac{T}{2} \log (1-q) - \frac{Tq}{2} \ .
\end{equation}
This function can be shown to coincide with the correct $V(q,T)$ on each stationary point of
$V(q,T)$.
As we are interested only in the value of $V(q,T)$ on its stationary points, the use of the
correct $V(q,T)$ calculated in Refs.~\cite{FP1,BFP} or
of the one given by Eq.~\ref{Vq} will give exactly the same result.

Note that, while the model is defined only for integer $p$, Eq.~\ref{Vq} holds also
for real $p$; we will therefore discuss the behavior of the different quantities for any real
$p\geq2$. In particular, the $p\rightarrow2$ limit is interesting being related to
a diverging fragility ($T_d\rightarrow T_K$) and to the discontinuous 1RSB transition
becoming a continuous one.

\subsection{Temperatures}

From Eq.~\ref{Vq} we can compute the three temperatures $T_K$, $T_g$ and $T_d$ as
functions of $p$. Their behavior is reported in Fig.~\ref{fig_2}. We immediately note that, for
$p\sim2$, the difference between $T_K$ and $T_g$ is very small, therefore the system is very
fragile; moreover, for $p\rightarrow\infty$
the Kauzmann temperature approaches zero (as $1/\sqrt{\log p}$),
while the glass transition temperature remains finite.
The system therefore becomes stronger and stronger on increasing $p$.

\begin{figure}[t]
\centering
%\vspace{.05cm}
\includegraphics[width=.45\textwidth,angle=0]{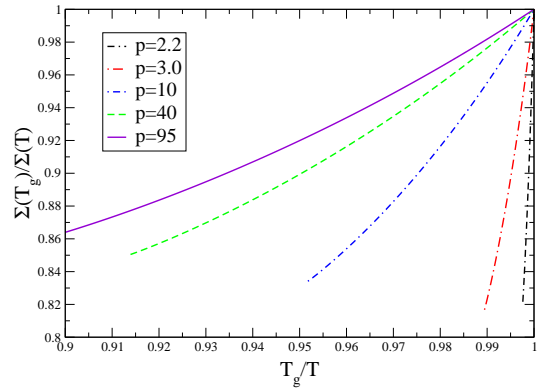}
\caption{The complexity $\Sigma(T_g)/\Sigma(T)$ as a function of $T_g/T$ for the p-spin
spherical model at different values of $p$.
Fragility is the slope of the curves in $T_g/T=1$.
The system becomes stronger on increasing $p$.}
\label{fig_3}
\end{figure}

\subsection{Complexity and fragility}

\begin{figure}[t]
\centering
%\vspace{.05cm}
\includegraphics[width=.45\textwidth,angle=0]{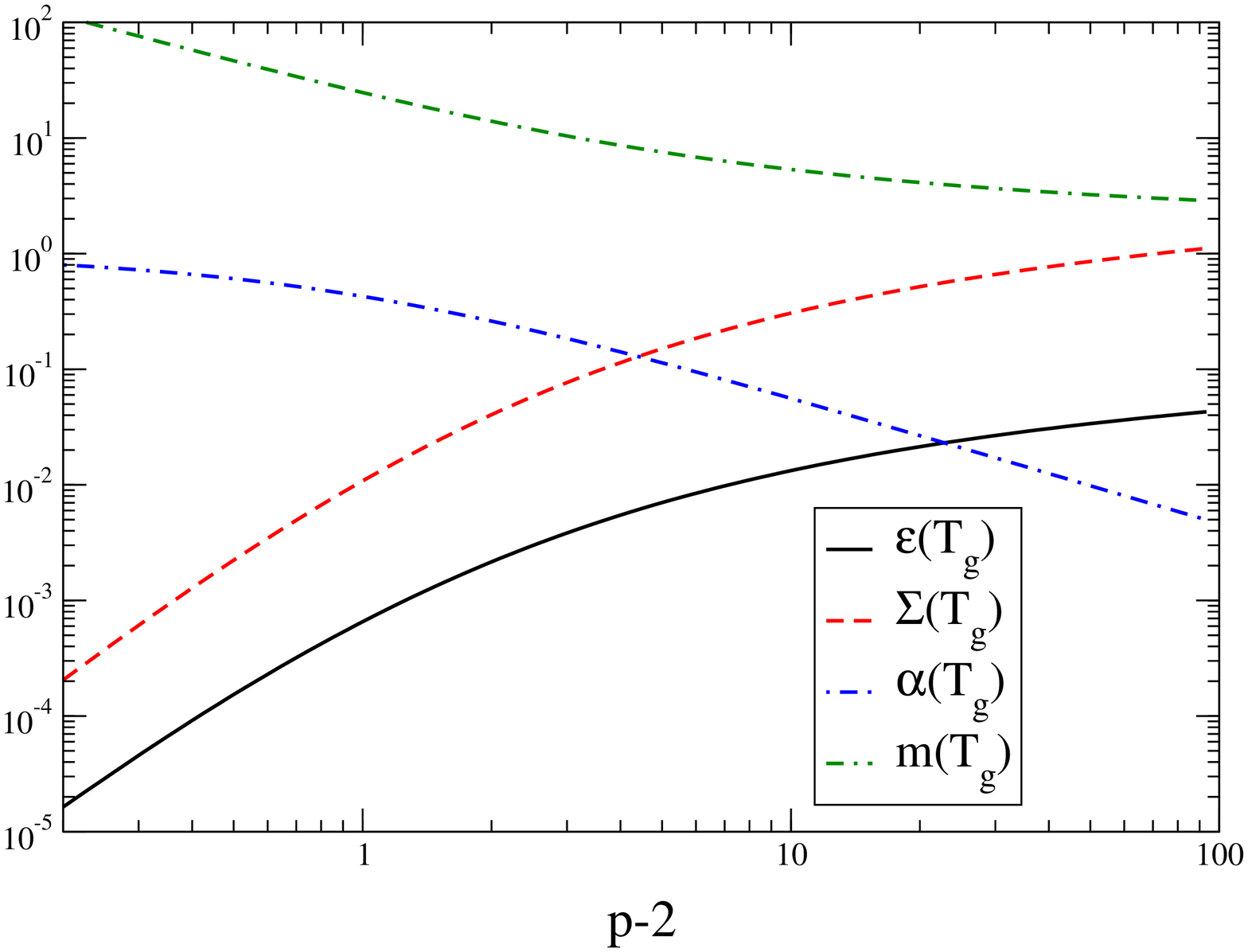}
\caption{Fragility $m(T_g)$, configurational entropy $\Sigma(T_g)$, ``volume'' of the equilibrium
states $\alpha(T_g)$ and barrier height ${\cal E}(T_g)$ for the p-spin spherical model as a
function of $p-2$.}
\label{fig_4}
\end{figure}

The same observation can be made more quantitative by considering an ``Angell plot'' for the
complexity \cite{Ang1}:
in Fig.~\ref{fig_3} we show the complexity $\Sigma(T)$ as a function of
the temperature, for different values of $p$. The choice of the particular scaling that appears
in Fig.~\ref{fig_3} has been made in order to make a close correspondence with Fig.~2 of
Ref.~\cite{Ang1}.
We see that the curves for different values of $p$ are ordered from bottom to top.
The same behavior is observed in experimental systems of different fragility. Indeed, the index
of fragility defined in Eq.~\ref{fragilita} is exactly one plus the slope of the curves
in $T_g/T=1$ (see Fig.~\ref{fig_3}):
\begin{equation}
\begin{split}
m(T_g)&=1+T_g \frac{\Sigma'(T_g)}{\Sigma(T_g)}  \\
&= 1 + \left. \frac{d [\Sigma(T_g)/\Sigma(T)]}{d [T_g/T]} \right|_{T=T_g} \ .
\end{split}
\end{equation}
The fragility index $m$ is shown in Fig.~\ref{fig_4} as a function of $p$. We see that it is
a decreasing function of $p$. Its values are in the range observed for experimental system due
to our (arbitrary) choice of the constant ${\cal C}$ appearing in Eq.~\ref{defTg}, ${\cal C}=0.1$.
In Fig.~\ref{fig_4} $\Sigma(T_g)$ is also reported as a function of $p$. We see that it is an
increasing function of $p$, that diverge as $\log p$ for $p \rightarrow \infty$:
thus, the number of states in this system is a decreasing function
of the fragility, at variance with what is currently believed (for a review, see Ref.~\cite{noifrag}).
We will discuss this point in detail in section~\ref{correlazioni}.

\subsection{Barrier heights and volume of the states}

In Fig.~\ref{fig_4} the barrier height ${\cal E}(T_g)$ is also reported as a function of $p$,
together with the index $\alpha(T_g)=1-q(T_g)$ that we called ``volume'' of the equilibrium
states at $T_g$. We observe that in this model the states become smaller on increasing $p$,
while the barriers separating them increase. 
In section~\ref{correlazioni} we will discuss this behavior trying to deduce a
geometric description of the evolution of the phase space of this model
at different $p$, and relate fragility to geometric properties of the phase space.

\section{Ising p-spin model}
\label{sec:ising}

The Ising p-spin model is another popular model for the study 
of the glass transition \cite{Derrida,KTW}.
Its Hamiltonian is given by Eq.~\ref{Hpspin}, where the variables $\sigma_i$ are Ising spins,
$\sigma_i = \pm 1$, and the spherical constraint is absent.
For the Ising p-spin model, the two-replica potential $V(q,T)$ is given by
\begin{equation}
\begin{split}
V(q,T)&=\beta \frac{p-1}{4} q^p + \beta \frac{p}{4} q^{p-1} \\
&- \frac{\int {\cal D}z \cosh(\Lambda z) \log \cosh (\Lambda z)}{\int {\cal D}z \cosh(\Lambda z)} \ ,
\end{split}
\end{equation}
where ${\cal D}z = \exp(-z^2/2) \ dz$, and $\Lambda^2 = \beta^2 \frac{p}{2} q^{p-1}$.

The Ising p-spin model is also a ``Gaussian-like'' model, like the spherical one.
However, the total number of states in the Ising p-spin model cannot be greater than $2^N$
(the total number of configurations), and hence $\Sigma(T) \leq \log 2$,
while in the spherical model $\Sigma(T_g)$
diverge as $\log p$ for $p \rightarrow \infty$, as previously discussed.

\subsection{Temperatures}

\begin{figure}[t]
\centering
%\vspace{.05cm}
\includegraphics[width=.45\textwidth,angle=0]{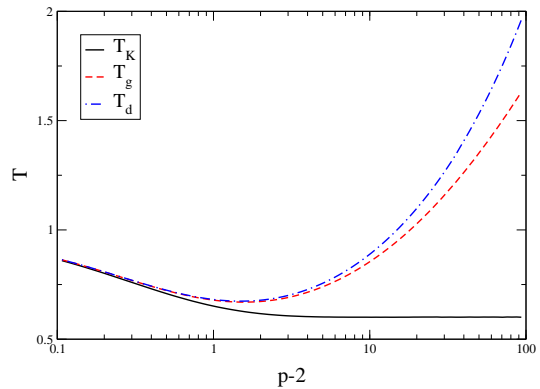}
\caption{Thermodynamic transition temperature $T_K$, glass transition temperature $T_g$
and dynamical transition temperature $T_d$ for the p-spin Ising model as a function
of $p-2$.}
\label{fig_5}
\end{figure}

The first consequence of this difference is observed when studying the transition temperatures
as a function of $p$ (see Fig.~\ref{fig_5}).
Indeed, as in the spherical model, we have $T_K \sim T_g$ for $p \sim 2$,
and $T_g \gg T_K$ for $p \rightarrow \infty$. But, in this model, $T_K$ tends to a finite value
at large $p$, while $T_g$ and $T_d$ diverge.
This behavior can be understood recalling that for a ``Gaussian-like'' model we have
$T_K \sim 1/\sqrt{\Sigma^*}$, $\Sigma^*$ being the total number of
states, i.e., the maximum of $\Sigma(T)$ \cite{noifrag}.

\subsection{Complexity and geometric properties of the phase space}

The ``Angell plot'' for the complexity of the Ising p-spin model looks very similar to the
one of the spherical model (see Fig.~\ref{fig_3}) and is not reported here.

\begin{figure}[t]
\centering
%\vspace{.05cm}
\includegraphics[width=.45\textwidth,angle=0]{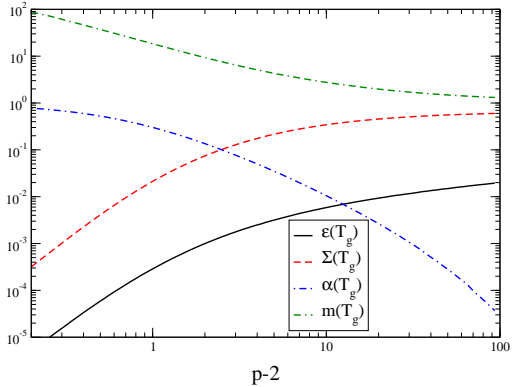}
\caption{Fragility $m(T_g)$, configurational entropy $\Sigma(T_g)$, ``volume'' of the equilibrium
states $\alpha(T_g)$ and barrier height ${\cal E}(T_g)$ for the p-spin Ising model as a
function of $p-2$.}
\label{fig_6}
\end{figure}

Having fixed an appropriate value for the constant ${\cal C}$ in Eq.~\ref{defTg} (${\cal C}=0.02$,
different from the value chosen in the previous case),
the behavior of the fragility as a function of $p$ is also very similar to the one of the
spherical model. The same behavior is found
for the other quantities under study, as one can deduce from
a comparison of Fig.~\ref{fig_6} and Fig.~\ref{fig_4}, the main difference being the discussed
behavior of $\Sigma(T_g)$ at large $p$.

\subsection{Vibrational properties and volume of the states}

Another relevant difference between the spherical and the Ising model is that, in the latter,
harmonic vibrations are not present (the variables being discrete): we have
$q(T) \rightarrow 1$ exponentially for $T \sim 0$, and the definition of $\alpha$
via Eq.~\ref{alphaTS}
gives $\alpha = 0$ for all $p$. However, the definition given in Eq.~\ref{alphaGP} and used in
our calculations gives a reasonable result also in absence of harmonic vibrations.

\section{Correlations between different properties of the phase space}
\label{correlazioni}

In this section we will examine the correlations in the quantities under study,
trying to relate fragility to the phase space geometry.
We will compare our results with the general consideration that we made in Ref.~\cite{noifrag},
and with the experimental results of Ref.~\cite{TS}.

\subsection{Fragility and volume of the states}

\begin{figure}[t]
\centering
%\vspace{.05cm}
\includegraphics[width=.45\textwidth,angle=0]{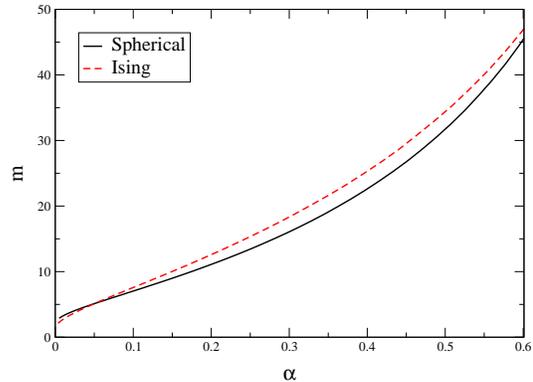}
\caption{Fragility versus $\alpha$ for the two investigated models. The curve is very similar
for the two models, and is consistent with the linear correlation found in \cite{TS} at least
for $\alpha \leq 0.4$.}
\label{fig_7}
\end{figure}

In Ref.~\cite{TS} it has been established that fragility is positively correlated with the index
$\alpha$ defined in section~\ref{sec:definitions}.
In other words, {\it fragile systems have large basins while strong systems have small basins}.
In Fig.~\ref{fig_7} we plot the fragility $m$ as a function of $\alpha$ parametrically in $p$
for the investigated systems. The curve $m(\alpha)$ is very similar for
the two models - remember that the only adjustable parameter is the constant ${\cal C}$
in Eq.~\ref{defTg}.
By comparison with Fig.~3 of
Ref.~\cite{TS}, we conclude that the model has a behavior similar to the one of real systems.
Surprisingly, also the linear correlation between $m$ and $\alpha$ is reproduced for
$\alpha \leq 0.4$.
Thus, mean field p-spin models are able to describe the relation between fragility and
the volume of the basins visited around $T_g$ found in Ref.~\cite{TS}.

\subsection{Fragility and total number of states}

\begin{figure}[t]
\centering
%\vspace{.05cm}
\includegraphics[width=.45\textwidth,angle=0]{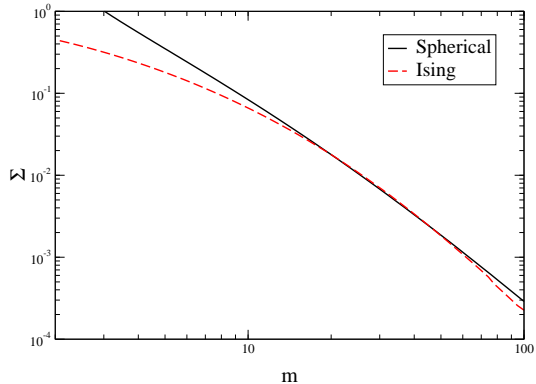}
\caption{Total number of states (represented by the complexity at $T_g$) as a function of
the fragility $m$: an inverse correlation is found between these quantities, at variance of
what is naively expected from the Adam-Gibbs relation.}
\label{fig_8}
\end{figure}

It is usually believed that fragile systems have a larger number of states than strong ones,
even if the total number of states is not an experimentally accessible quantity and numerical
simulations give contradictory results~\cite{numeric}.
However, in the model considered here the behavior is exactly the opposite. In Fig.~\ref{fig_8}
we report $\Sigma(T_g)$ as a function of the fragility: we see that the total number of states
is a decreasing function of the fragility, at variance of what is currently believed.
To discuss this point, we have to refer to \cite{noifrag}: there, we discussed the possibility
of correlating fragility with the total number of states for general models of $\Sigma(T)$, and
assuming the validity of the Adam-Gibbs relation, Eq.~\ref{AG}.
We concluded that the knowledge of the distribution of states is not enough to determine
the fragility. Indeed, the relevant
parameter was identified, for a general ``Gaussian-like'' distribution of states, as
\begin{equation}
\label{D}
D = \frac{{\cal E}(T_g)}{T_K \Sigma(T_g)} \ .
\end{equation}
Note that in Eq.~\ref{D} we have to calculate ${\cal E}$ at $T=T_g$ because in the considered models
the barrier height ${\cal E}$ is a $T$-dependent quantity, while in the Adam-Gibbs relation
it is usually assumed to be a constant (see Eq.~\ref{AG}). However, the Adam-Gibbs relation
has been tested around $T_g$, therefore, to a good approximation, we can fix ${\cal E}$ to be a
constant equal to its $T=T_g$ value.
The parameter $D$ is inversely proportional to the fragility $m$: therefore $m \sim \Sigma/{\cal E}$.
Thus, fragility is not simply correlated to the total number of states:
if the ``barrier heights'' grow faster than the total number of states,
fragility can be a decreasing function of $\Sigma$.
We will now show that this is indeed the case in the considered models.

\subsection{Barrier heights, total number of states and fragility}

From Fig.~\ref{fig_4} and Fig.~\ref{fig_6} we see that the barrier height is indeed an
increasing function of $p$ in the considered models.
Using Eq.~\ref{defTg}, Eq.~\ref{D} can be written as
\begin{equation}
D = {\cal C} \frac{T_g}{T_K} \ .
\end{equation}
Therefore, from Fig.~\ref{fig_2} and Fig.~\ref{fig_5} we see that $D$ is indeed an increasing
function of $p$ that diverge for $p \rightarrow \infty$, as the ratio $T_g/T_K$ increase on
increasing $p$ for both models.
Thus, we can conclude that in the considered models the height of the barriers (in units of $T_K$)
increases faster than the total number of states. This explains why one observe an inverse
correlation between fragility and the total number of states, as discussed above and in
Ref.~\cite{noifrag}.

\subsection{A geometric picture of the phase space}

\begin{figure}[t]
\centering
%\vspace{.05cm}
\includegraphics[width=.4\textwidth,angle=0]{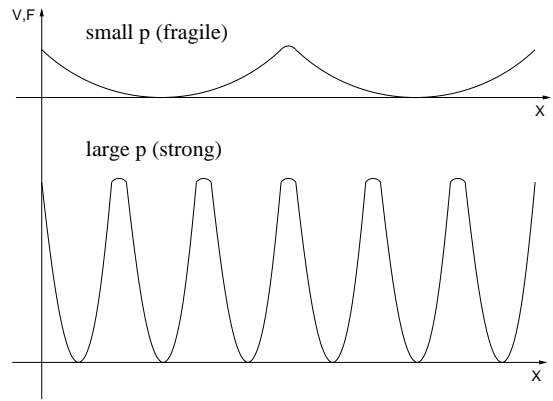}
\caption{Sketch of the evolution of the p-spin free energy by varying $p$: at small $p$
there is a small number of states of large volume separated by low barriers; at high $p$
there is a large number of states of small volume separated by high barriers. The height
of the barriers increase faster than the number of states: thus, fragility is a decreasing
function of $p$.}
\label{fig_9}
\end{figure}

Collecting all the information that we obtained in the previous sections, we can propose
a geometric picture of the variation with $p$ of the $p$-spin model free energy landscape.
Indeed, on increasing $p$: \\
{\it i)} The total number of states increases. \\
{\it ii)} The volume of the states decreases ($\alpha$ decreases). \\
{\it iii)} The height of the barriers between states increases. \\
Thus, we get the picture of a landscape where, on increasing $p$, a great number of small
states with very high curvatures and separated by very high barriers appear: a sketch of this
evolution is given in Fig.~\ref{fig_9}.
The behavior of the fragility in this situation is related to the behavior of $\Sigma/{\cal E}$,
the ratio between number of states and height of the barriers between them: in these models,
it turns out that ${\cal E}$ increase faster than $\Sigma$, and the fragility is a decreasing
function of $p$.

This behavior is consistent with the fact that fragility turns out to be
positively correlated with the ``volume'' of the states as measured by $\alpha$. Indeed,
if, on the contrary, the barrier height grew slower than the total number of states
(equivalently, if $m$ would be positively correlated with the total number of states),
there should be also an inverse correlation between $m$ and $\alpha$, in disagreement with what
is experimentally observed.

In the $p\rightarrow 2$ limit, where the fragility becomes infinite, the second
derivative with respect to $q$ of the potential $V(q,T)$ calculated in
$q=0$ and $T=T_K=T_d$ vanishes (see Fig.~\ref{fig_1}) 
and the so-called spin glass susceptibility
diverges at the critical temperature. In other words when the fragility becomes
infinite soft modes appear at the critical temperature supporting the
previously presented physical picture.

Note that the outlined picture is valid for ``Gaussian-like'' models, i.e., models where
the complexity is a concave function of the temperature that vanish at $T_K$ without any
inflection point. These models seem to describe correctly the distribution of basins in real
systems only for relatively high fragilities. The behavior of the complexity (or configurational
entropy, or excess entropy) as a function of temperature for very strong systems is still an
open problem; our discussion may not apply to these systems.

\section{Conclusions}

From the investigation of two mean field p-spin models, we proposed a
picture for the evolution of the free-energy landscape from fragile liquids to strong ones.
This picture accounts for the recently observed correlation between fragility of a liquid
and vibrational properties of the corresponding glass.
The main prediction of our analysis is that the total number of states and the Adam-Gibbs
parameter ${\cal E}$ should both be decreasing functions of the fragility. Unfortunately,
existing data are not sufficient to strictly test this prediction; the excess entropy
is available only for few experimental systems, and numerical simulations are performed
in a temperature range where the fragility of the investigated systems is approximately
the same. We hope that this predictions can be tested in the future.

We thank Stefano Mossa for a careful reading of the manuscript, and Cristiano De Michele,
Tullio Scopigno, Francesco Sciortino and Luca Angelani for useful discussions.

\appendix

\section{The two-replica potential}
\label{appVq}

\noindent
The two-replica potential is defined in Ref.~\cite{FP1} as
\begin{equation}
\begin{split}
&V(q)=-F(T)-\frac{T}{N} \int d\sigma \frac{e^{-\beta H(\sigma)}}{Z(\beta)} \log Z(\sigma,q) \\
&Z(\sigma,q)=\int d\tau e^{-\beta H(\tau)} \delta(q-q(\sigma,\tau)) \ ,
\end{split}
\end{equation}
where $F(T)$ is the equilibrium free energy, and
$q(\sigma,\tau)$ is the overlap function.
The following expression is then derived \footnote{Note that the expression given in
\cite{FP1}, Eq. (15), is slightly different from this one probably due to a typing mistake}:
\begin{equation}
\begin{split}
&V(q) = - F(T) \\ &- \lim_{n\rightarrow0} \lim_{m\rightarrow 1} \frac{T}{Nn}
\frac{\partial}{\partial m}  \left( \int d\sigma e^{-\beta H(\sigma)} Z(\sigma,q)^{m-1} \right)^n \ .
\end{split}
\end{equation}
The last integral can be rewritten as
\begin{equation}
\begin{split}
&\left( \int d\sigma e^{-\beta H(\sigma)} Z(\sigma,q)^{m-1} \right)^n = \\
&\int d\sigma_{a \alpha} e^{-\beta \sum_{a \alpha} H(\sigma_{a\alpha})}
\prod_{a=1}^n \prod_{\alpha=2}^m \delta(q-q(\sigma_{a1},\sigma_{a\alpha})) \ ,
\end{split}
\end{equation}
where $a=1,\cdots,n$, $\alpha=1,\cdots,m$. This is exactly the expression
of the $nm$ times replicated equilibrium partition function, with the additional
constraint given by the $\delta$-functions.
Using standard manipulations \cite{CS}, it is rewritten as
\begin{equation}
\begin{split}
&\int d Q_{a\alpha,b\beta} \ e^{N f(Q)} \ \prod_{a=1}^n \prod_{\alpha=2}^m \delta(q-Q_{a1,a\alpha}) \\
&f(Q)=\frac{\beta^2}{4} \sum_{a\alpha,b \beta} Q_{a\alpha, b \beta}^p + \frac{1}{2} \log \det Q \ .
\end{split}
\end{equation}
Thus, evaluating the integral at the saddle point, we get
\begin{equation}
\label{Vq1}
V(q)=-F(T) - \lim_{n\rightarrow0} \lim_{m\rightarrow 1} \frac{T}{n}
\frac{\partial}{\partial m} f(\bar{Q}) \ .
\end{equation}
The matrix $\bar{Q}$ is defined by the following conditions: \\
{\it i)} the elements on the diagonal are equal to 1; \\
{\it ii)} the elements $\bar{Q}_{a1a\alpha}$, $\alpha>1$, are equal to $q$; \\
{\it iii)} all the other elements are determined by the maximization of $f(Q)$. \\
As usual, one needs a parametrization of the matrix $Q$ in order to perform the analytic
continuation to non-integer $n$ and $m$. A possible {\it ansatz} is \cite{FP1} (in the example,
$n=3$, $m=4$):
\begin{equation}
\nonumber
\bar{Q}=\left(
\begin{array}{ccc}
\left(
\begin{array}{cccc}
1 & q & q & q \\
q & 1 & r & r \\
q & r & 1 & r \\
q & r & r & 1 \\
\end{array}
\right) & 0 & 0 \\
0 & \left(
\begin{array}{cccc}
1 & q & q & q \\
q & 1 & r & r \\
q & r & 1 & r \\
q & r & r & 1 \\
\end{array}
\right) & 0 \\
0 & 0 & \left(
\begin{array}{cccc}
1 & q & q & q \\
q & 1 & r & r \\
q & r & 1 & r \\
q & r & r & 1 \\
\end{array}
\right) \\
\end{array}
\right)
\end{equation}
Within this {\it ansatz}, and using the relation
\begin{equation}
\det  \left(
\begin{array}{cccc}
1 & q & q & q \\
q & 1 & r & r \\
q & r & 1 & r \\
q & r & r & 1 \\
\end{array}
\right) = (1-r)^{m-2} [ 1-2r+rm-(m-1)q^2 ] \ ,
\end{equation}
one gets
\begin{equation}
V(q)=-\frac{\beta q^p}{2} + \frac{\beta r^p}{4}
 - \frac{T}{2} \left[ \log(1-r) + \frac{r-q^2}{1-r} \right] \ ,
\end{equation}
where $r(q)$ is determined by $\partial_r V=0$.
Now, it is easy to check that the condition $dV/dq=\partial_q V=0$, together with
$\partial_r V=0$, is satisfied if $q=r$. Thus, when $V(q)$ is stationary, $r(q)=q$ and
the potential $V(q)$ reduces to the one given by Eq.~\ref{Vq}. \\
The fact that when $dV/dq=0$ the matrix $\bar{Q}$ reduces to the usual 1RSB overlap matrix
(let us call it ${\cal Q}$) is general: indeed, the condition $dV/dq$ from Eq.~\ref{Vq1}
is equivalent to
\begin{equation}
\frac{df(Q)}{dQ}=0 \ .
\end{equation}
This means that the function $f(Q)$ must be stationary with respect to all the elements of $Q$
if $dV/dq=0$,
and we know that the 1RSB matrix ${\cal Q}$ provides a solution to this condition.
As a final remark, we note that if $\bar{Q}={\cal Q}$, we have
\begin{equation}
\frac{f(\bar{Q})}{nm} = -\beta \phi_{1RSB}(m,q) \ ,
\end{equation}
where $\phi_{1RSB}$ is the usual 1RSB free energy. Substituting this expression in Eq.~\ref{Vq1},
one obtains
\begin{equation}
\begin{split}
V(q)&=-F(T)+\lim_{m\rightarrow1} \partial_m \Big( m \phi_{1RSB}(m,q) \Big) \\
&= \lim_{m\rightarrow1} \partial_m \phi_{1RSB}(m,q) \ ;
\end{split}
\end{equation}
using the relation $\phi_{1RSB}(m=1)=F(T)$ that holds above $T_K$.
Therefore, {\it on its stationary points}, $V(q)$ is given (at the 1RSB level)
by this simple expression, that can be easily calculated in several models.
Note that, as discussed in Ref.~\cite{BFP}, full RSB effects can be important for the computation
of $V(q)$. However, we don't account for them in this paper.

%%%%%%%%%%%%%%%%%%%%%%%%%%%%%%%%%%%%%%%%%%%%%%%%%%%%%%%%%%%%%%%%%%%%%%%%%%%
%                             REFERENCES
%%%%%%%%%%%%%%%%%%%%%%%%%%%%%%%%%%%%%%%%%%%%%%%%%%%%%%%%%%%%%%%%%%%%%%%%%%%

\end{document}